\documentclass[pre, aps,showpacs,twocolumn,preprintnumbers,amsmath,amssymb,superscriptaddress]{revtex4-1}

\usepackage{graphicx}

\usepackage{dcolumn}

\usepackage{bm}
\usepackage{multirow}
\usepackage{tabularx}

\begin{document}

\title{Mesoscopic Model for Free Energy Landscape Analysis of DNA sequences.}

\author{R. Tapia-Rojo}

\affiliation{Dpto. de F\'{\i}sica de la Materia Condensada,
Universidad de Zaragoza. 50009 Zaragoza, Spain}

\affiliation{Instituto de Biocomputaci\'on y F\'{\i}sica de Sistemas
Complejos, Universidad de Zaragoza. 50009 Zaragoza, Spain}

\author{D. Prada-Gracia}

\affiliation{Freiburg Institute for Advanced Studies (FRIAS), University of Freiburg, 79104 Freiburg, Germany}

\author{J.~J. Mazo}

\affiliation{Dpto. de F\'{\i}sica de la Materia Condensada,
Universidad de Zaragoza. 50009 Zaragoza, Spain}

\affiliation{Instituto de Ciencia de Materiales de Arag\'on,
C.S.I.C.-Universidad de Zaragoza. 50009 Zaragoza, Spain.}

\author{F. Falo}

\affiliation{Dpto. de F\'{\i}sica de la Materia Condensada,
Universidad de Zaragoza. 50009 Zaragoza, Spain}

\affiliation{Instituto de Biocomputaci\'on y F\'{\i}sica de Sistemas
Complejos, Universidad de Zaragoza. 50009 Zaragoza, Spain}


\begin{abstract}

A mesoscopic model which allows us to identify and quantify the strength
of binding sites in DNA sequences is proposed. The model is based on
the Peyrard-Bishop-Dauxois model for the DNA chain coupled to a
Brownian particle which explores the sequence 
 interacting more importantly with open base pairs of the DNA chain. 
We apply the model to promoter sequences of different organisms. The free energy landscape
obtained for these promoters shows a complex structure that is
strongly connected to their biological behavior. The analysis method
used is able to quantify free energy differences of sites within
genome sequences.

\end{abstract}

\maketitle

\section{Introduction}

The study of biomolecules at a mesoscopic level tries to identify the
main degrees of freedom of the system and understand its behavior in
terms of the dynamical and statistical-mechanics properties of the
model. At this level, the concept of a free-energy landscape (FEL)
represents a paradigm for the comprehension of several biological
complex problems such as protein folding, protein structure, and
biomolecular interaction ~\cite{Wales}. A FEL gives the change in the free
energy of a system when the different degrees of freedom change.  The
description given by the topology of the FEL permits to connect structure, 
dynamics, and thermodynamics in many different systems ranging
from atomic clusters to biomolecules or soft matter systems ~\cite{Wales2}.


Here we address our attention to the characterization of the FEL of
DNA sequences and, in particular, those sites with regulatory and
transcriptional relevance. Recently, a great attention has been
devoted to the mechanism whereby proteins bind to specific sites on
DNA~\cite{search}. The quantification, grounded in a physical basis, of the
strength of these binding sites is an open problem. In this paper, we
propose a model which allows us to calculate free energy differences
between specific and nonspecific binding sites. Even more, we are
able to build, from the trajectories of the model, a representation of
the free energy landscape.

The model is inspired in the protein search of the binding sites in a
DNA chain. This search is a combination of three-dimensional (3D) jumps between separated
regions of DNA and one-dimensional (1D) diffusion along the chain. It has been stated
that most of the search time is spent in the 1D diffusion process
since the time jumps in three dimensions is negligible \cite{PhysicsMolBio, Berg,
  Marko}. Thus, the restriction to a 1D search is a good starting point
for our model.  On the other hand, it has also been conjectured that
the dynamics of the DNA chain plays an important role in the
recognition of binding sites by the regulatory factors or the
transcription protein \cite{Kalosakas04, Apostolaki11}. Thus,
transcription processes, for instance, would be induced by the binding
of RNA-polymerase to openings (bubbles) in the DNA chain. This idea is
supported by computational approaches to the DNA dynamics
\cite{Alexandrov_PLoSCB}, and experimental evidences
\cite{expDNA}. Our model follows this idea and considers the
interaction of a test particle, which explores the DNA chain, coupled
to the bubbles. In order to be physically and biologically relevant,
such a description should provide useful qualitative and quantitative
information about the process. A similar strategy has been used to
characterize complex networks and identify regions of special
relevance (communities). In such an approach, a ``fictitious" Brownian
particle goes over the graph \cite{PNASnet} and its dynamics reveals
``thermodynamics" and structural quantities of the topology \cite{JGG}.


In this paper, we propose a mesoscopic model for DNA-particle interaction. In our
picture, the test particle undergoes a 1D Brownian motion in
interaction with a classical field, the DNA chain itself, whose
dynamics is also affected by the presence of the particle.
The test particle interacts more strongly with open base-pairs of the DNA chain. 
In this way ``softer" regions of the DNA sequence are more likely to be visited by the particle, 
which will help also in stabilizing the bubbles. This interaction is not sequence-dependent, as the
DNA base-pair dynamics already depends on the sequence. Thus, this model could also represent
the interaction of a real protein as RNA-polymerase, with the DNA bubbles.

Particle and chain are described at the same level of complexity. We use the
Peyrard-Bishop-Dauxois (PBD) \cite{PBD,Peyrard_NonL} model to perform
the dynamics of the chain. This model was proposed initially for the
study of DNA thermal denaturation and incorporates the formation and
dynamics of bubbles in a natural way. The PBD model can take into account the sequence
information through its parameters.

Our model incorporates three basic ingredients of the physics of the
system: DNA sequence, bubble dynamics, and 1D particle diffusion. The
analysis of the DNA-particle complex allows us not only to identify
possible binding sites, but also to describe the whole structure of
the free-energy landscape and determine free-energy differences
between different representative states. We show the validity and
usefulness of our approach by studying the FEL of three promoter
sequences. In each case, the FEL topology gives insight into the biological
behavior of the system.

\section{Model}
We describe the DNA chain by a modified
Peyrard-Bishop-Dauxois model\cite{PBD,Peyrard_NonL,RTR10}. There, the
complexity of the DNA molecule is reduced to the study of the dynamics
of $N$ base pairs described by the variables $y_n$ defining the
distance between the bases. In this framework $H_{DNA}=\sum_{n=1}^N \left[
  \frac{p_n^2}{2m}+V(y_n)+W(y_n,y_{n-1}) \right]$. In our model
\begin{equation}
V(y)=D ({\rm e}^{-\alpha y}-1)^2 + G {\rm e}^{-(y-y_0)^2/b},
\label{eq:V}
\end{equation}
where the first term accounts for the hydrogen bond interaction and
the second one for interactions with the solvent \cite{Weber,RTR10}.
It was shown in \cite{RTR10} that the inclusion of this barrier
modifies drastically the duration and stability of the bubbles.

\begin{figure}
\includegraphics[width=8.0 cm ]{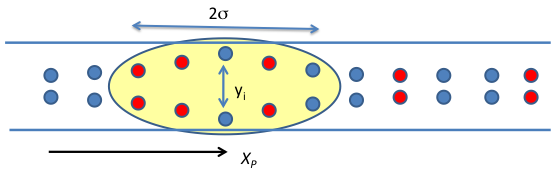}
\caption{(Color online) Schematic illustration of the DNA-particle interaction model. The unidimensional chain (solid circles) is represented by the base pair opening coordinate ($y_i$) while the probe particle (shaded ellipse) is a diffusing particle along the DNA chain ($X_p$).}
\label{fig:model}
\end{figure}

The potential $W(y_n,y_{n-1})$ describes the stacking interaction between the base pairs along the DNA strand,

\begin{equation}
W(y_n,y_{n-1})=\frac{1}{2}K\left(1+\rho e^{-\delta(y_n+y_{n-1})}\right)(y_n-y_{n-1})^2.
\end{equation}

In order to study different DNA sequences, the PBD model includes
sequence dependent Morse parameters: $D_n$, $\alpha_n$. Regarding the
DNA chain, we  use the set of parameters~\cite{param} considered
in~\cite{RTR10}.

The particle is represented by a Brownian particle (see
Fig. \ref{fig:model}) moving in a one-dimensional space with
coordinate $X_p$ and interacting with the DNA chain through a
phenomenological potential which depends on $X_p$ and the set of
coordinates $\{y_i\}_{i=1}^N$: $H_P=p_p^2/2m_p+V_{int}(X_p, \{y_i\})$
with

\begin{equation}
V_{int}(X_p, \{y_i\}) = -\frac{B}{\sqrt{\pi\sigma^2}}  \sum_{i} \tanh (\gamma y_i ) \ e^{-(X_p-ia)^2/\sigma^2}
\label{eq:Vint}
\end{equation}
where $B$ sets the interaction amplitude, $\gamma$ the range of
interaction with the base separation and $\sigma$ the spatial range of
interaction on the DNA chain. The functional form for the interaction
has been chosen to be linear at low $y_i$ and to saturate at large $y_i$
in order to avoid that the chain opens indefinitely. Note that with
this term the particle is trying to open the chain in a length range
of $\sigma$ and get self-trapped. Although possible, no sequence dependence is included in this term since we are interested in giving a general picture of the
FEL.


We still have to fix the parameters for this interaction term. For the
particle damping and mass, we  take $\eta_p=10^{14}s^{-1}$ and
$m_P=7000Da$. These values are of the order of magnitude of proteins
which bind DNA \cite{hop, biol}. The intensity of the interaction
chosen is $B=0.52 eV$. This value provides local interactions of the
order of the Morse potential dissociation energy at each base
pair. The parameter $\gamma=0.8\AA^{-1}$ saturates the interaction at
$y=1.25\AA$, a typical value for open base pairs. We take $a=1$ for
the longitudinal separation between base pairs, in arbitrary units,
and consider $\sigma=3$. This provides an interaction range of around
$5$ or $6$ base pairs (bp). It is interesting to note that this value has been
chosen in order to observe states with bubbles of $10-20$ bp, which is an adequate width for the processes we take into
account here \cite{arkiv}.

Once we have fixed the model parameters, we derive the Langevin
equations for both the chain bases and the particle. For the chain, we get:
\begin{eqnarray} 
m\frac{\partial^2 y_n}{\partial t^2}+m\eta\frac{\partial y_n}{\partial t} &=&-\frac{\partial\left [W(y_n,y_{n+1}+W(y_{n-1},y_m)\right]}{\partial y_n}  \nonumber \\ & & -\frac{\partial V}{\partial y_n}-\frac{\partial V_{int}}{\partial y_n}+\xi_n(t),
\end{eqnarray}
where $\eta$ stands for the damping and $\xi_n$ for the thermal noise, so $\langle \xi_n(t)\rangle=0$ and $\langle \xi_n(t)\xi_k(t')\rangle=2m\eta k_BT\delta_{nk}\delta(t-t')$ hold.

The Langevin equation for the particle is:
\begin{equation}
m_p\frac{\partial^2 X_p}{\partial t^2}+m_p\eta_p\frac{\partial X_p}{\partial t}=-\frac{\partial V_{int}}{\partial X_p}+\xi_p(t).
\end{equation}
where, $\eta_p$ stands for the particle damping and $\xi_p$ for the thermal noise. The fluctuation-dissipation relation reads as $\langle \xi_p(t)\rangle=0$ and $\langle \xi_p(t)\xi_p(t')\rangle=2m_p\eta_p k_BT\delta(t-t')$.

\begin{figure}
\centering\includegraphics[width= 11.0 cm ]{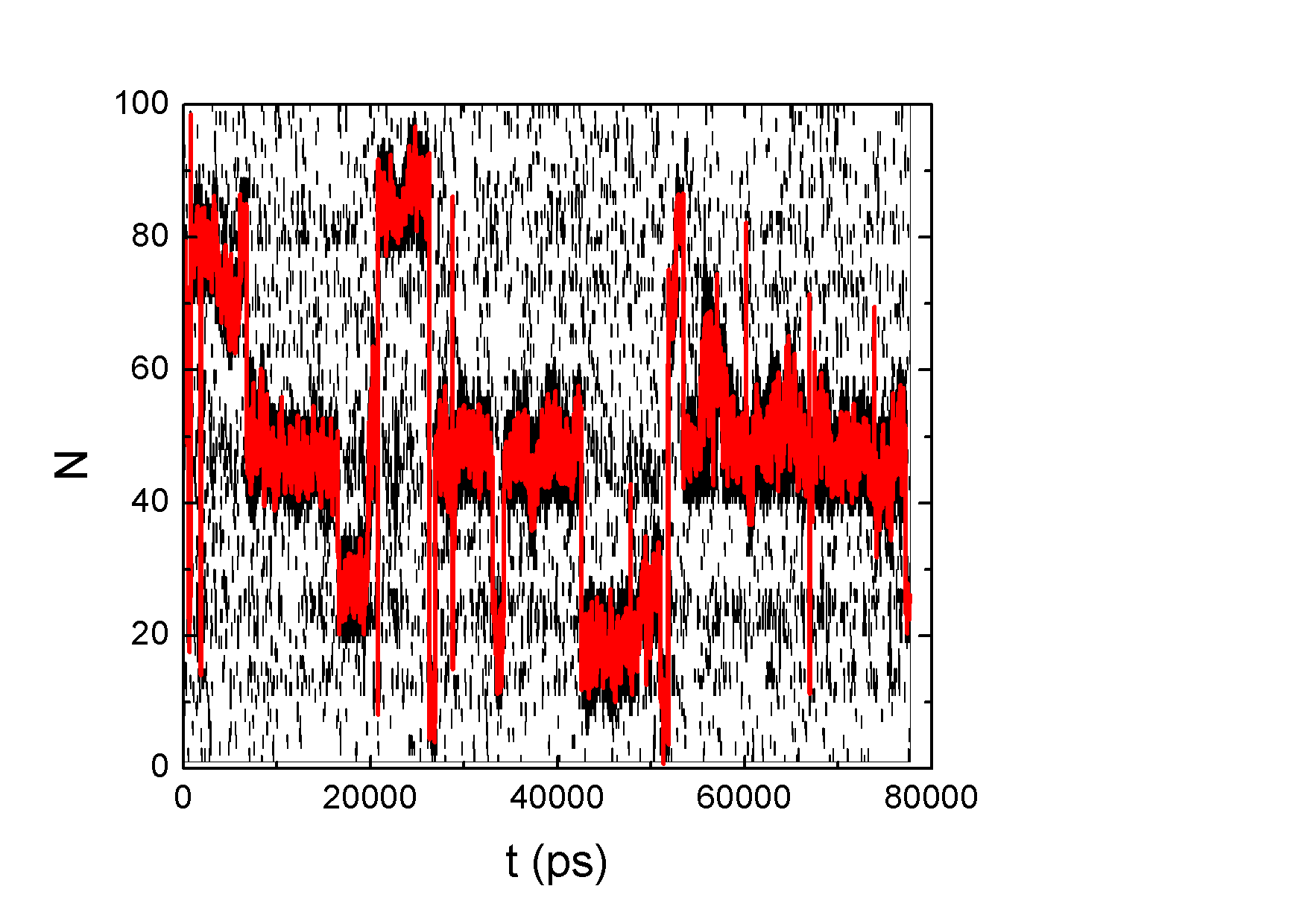}
\caption{(Color online) Trajectory of the particle shown in red line (light gray) with the DNA chain for a collagen sequence (see results section). Black points stands for open bp ($y_{AT} > 1.\AA$  or $y_{CG} > 1.5\AA$) and white for closed ones.}
\label{fig:tray} 
\end{figure}

The equations were numerically integrated using the stochastic
Runge-Kutta algorithm \cite{sde2}.  The integration of the Langevin
equations of motion provides trajectories of the particle and the DNA
chain. Each DNA sequence was simulated in five different realizations
for $40\mu s$, using $10$ fs time steps and a $1\mu s$ preheating
time. Since it has been reported that 1D diffusion periods cover a
time of the order of milliseconds \cite{hop, biol}, the simulation time used is
reasonable for the problem considered here. The simulation temperature
is $T=290K$ and the boundary conditions for the protein are periodic,
while for the chain we consider the hard boundary conditions
discussed in \cite{RTR10}.  An example of such a trajectory is
given in Fig. \ref{fig:tray}. It is observed that the particle moves in a
``sea'' of open bubbles which clearly shows the soft
domains of  the genome structure. 
The particle eventually jumps between these domains opening large stable bubbles. 
As seen in the figure the dynamics of the bases is strongly affected by the presence
of the particle.

\begin{figure*}
\centering\includegraphics[width=0.8\textwidth ]{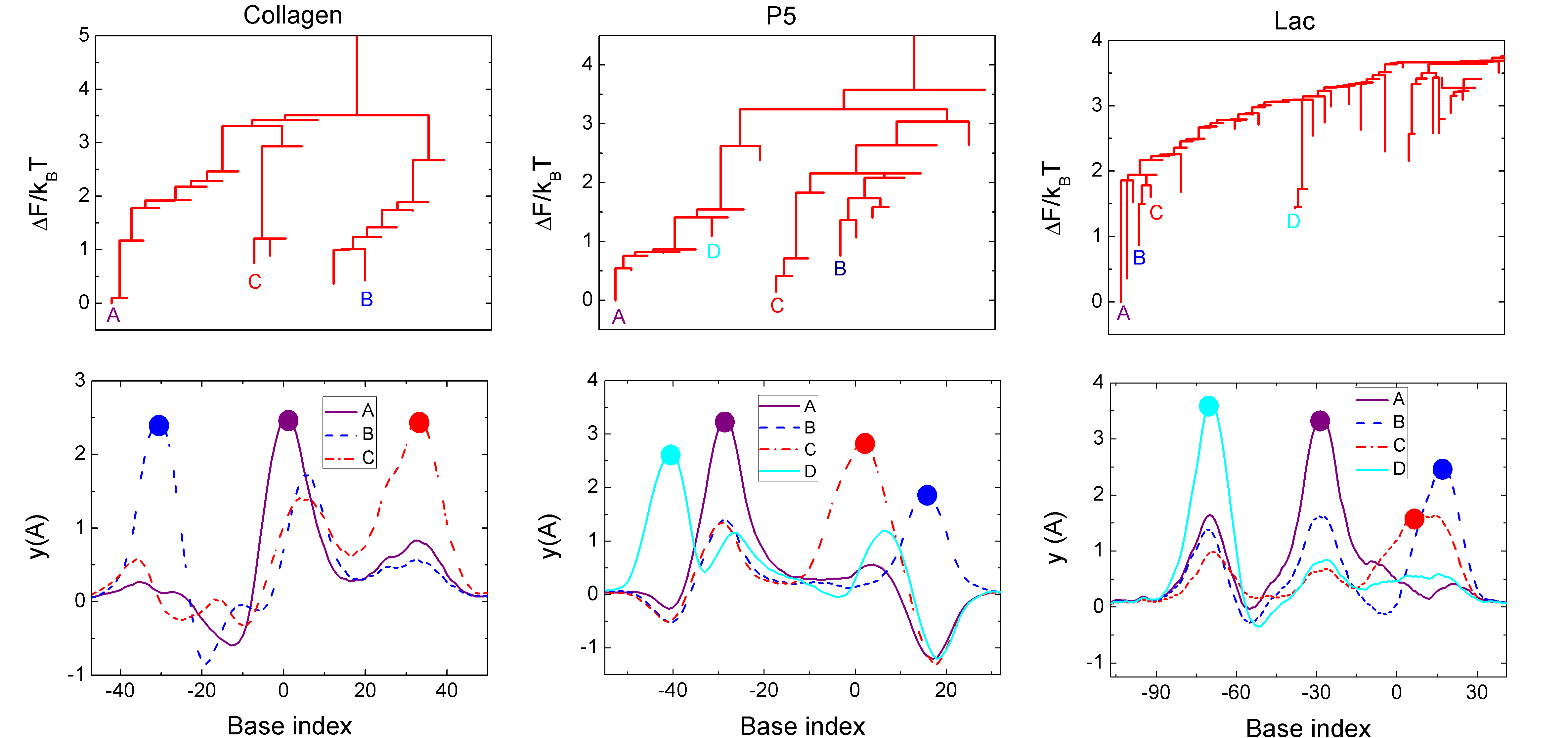}
\caption{(Color online) Free energy dendrograms (top) for each of the three
  sequences together with selected states (bottom). The
  dendrograms represented are a detail of the whole structure showing
  low energy basins only. The construction of the dendrogram has been performed considering the representative node of each basin. Significant biological states have been
  searched within the network structure. They correspond to the
  important basins of the hierarchical free energy organization.}
\label{fig:dendos} 
\end{figure*}

\section{Analysis}

To extract useful information of such trajectories, and due to the
large dimensionality of the system, we apply the principal component
analysis (PCA) \cite{jolliffe_book} to the chain trajectory. It has
been proved \cite{RTR10} that the first few eigenvectors unveil the
softest regions of the DNA chain and hence the possible binding sites
for our particle. Even more, PCA reduces the large
number of degrees of freedom of the system to just a few, by
projecting the coordinates of the system into the first few
eigenspaces (reduced trajectories). For each of the sequences
considered, we  restrict ourselves to the first five
eigenspaces. This subspace accounts for  $75\%$ of the total
fluctuations of the chain dynamics.

To obtain the FEL properties of the system we make use of the map
of trajectories to a conformational Markov network (CMN) [17,20]. The
CMN has been proven to be a useful representation of large stochastic
trajectories \cite{rao,caflisch,gfeller}.  This coarse grained picture
is usually constructed by discretizing the conformational space
explored by the dynamical system and considering the hops between the
different configurations as dictated by the MD simulation. In this
way, the nodes of a CMN are the subsets of configurations defined by
the conformational space discretization, and the links between nodes
account for the observed transitions between them. The information of
the stochastic trajectory allows us to assign probabilities for the
occupation of a node ($P_i$) and for the transitions between two
different configurations ($P_{ij}$). Defined as above, a CMN is thus a
weighted and directed graph. It should be stressed that the
information contained in the CMN is much richer that one given by
equilibrium statistical mechanics since it includes the {\it dynamics} of
the system encoded in the probability transitions, $P_{ij}$.

In our case, we start from the reduced trajectory for the DNA
(obtained using five principal components) and the trajectory of the
test particle. We discretize the total coordinate space in 20 bins
of equal volume for the reduced trajectory and $N$ bins (the DNA base pairs) for
the particle. This will constitute the microstate space of the CMN,
each node with occupancy probability $P_i$ obtained from the reduced
trajectory. Once the CMN has been built, we split it into basins of
attraction, \emph{i.e.}, regions in which the probability fluxes
($P_{ij}$) converge to a common state (attractor) of the network. This
task is usually hard, since algorithms scale as power law of the
system size. In this case we have applied the stochastic steepest
descent algorithm developed in \cite{PLoSDPG}, which scale as $N \log
N$. In this decomposition, a basin corresponds to a coarse-grained state (of connected nodes) of the CMN. 
In next section, we  represent each basin by its attracting node.

Once these basins have been defined we can represent the FEL by a
hierarchical tree diagram (dendrogram) \cite{PLoSDPG}, built according
to the weights and links among the basins. This representation is
similar to the ``disconnectivity graph" scheme used in other context
\cite{Wales, Wales2, krivov}. First, an ``adimensional free energy" is
assigned to each node $i$ given by $F_{i}/kT=log(P_{w})-log(P_{i})$
where $w$ represents the weightiest node. Using this magnitude as
control parameter, we slowly increase it step by step from its zero
initial value.  At each step of this process, we obtain a network
composed of those nodes with free energy lower than the current
threshold value. As the free-energy threshold increases, new nodes
emerge together with their links. These new nodes may be attached to
any of the nodes already present in the network or they can emerge as
a disconnected component. At a certain value of $F/kT$, some
components of the network become connected by the links of a new node
incorporated at this step. Initially we have a set of disconnected
vertical lines (corresponding to basins) which become linked once the control
parameter has overcome the barriers between them {\emph i. e.} when 
the free energy of the saddle nodes is reached. Then we draw a horizontal line 
linking these two basins. Obviously, for large
threshold all the network is connected. We can plot this process as a
``tree diagram" or dendrogram.

Using this representation we can understand qualitatively and
quantitatively the hierarchical organization of the basins and the
barriers among them and figure out the behavior of different
sequences.

This method could be applied to a DNA chain without a particle. However, 
the inclusion of the particle is essential to get the FEL of the
system. An analysis of the DNA alone (PBD model) lets us  determine
the opening probabilities and average position of the chain base
pairs, and unveils the softer regions that can indeed be related to
sites of biological importance. Nevertheless, the FEL of this model is
trivial, as opening events are rare and the chain remains closed for
most of the time. The inclusion of the particle stabilizes the bubbles (as can be
observed in Fig. \ref{fig:tray})
and allows us to go further in terms of predictions. We are able
to define relevant states in a precise and systematic way (basins),
to predict possible binding sites, and to extract the thermodynamical
magnitudes related with them, thus characterizing  these sites in terms
of biological importance.

\section{Results}

To illustrate the method and validate our model, we  analyze three
different promoter sequences. Promoters are DNA regions in which
regulation and initiation of transcription of a gene occurs. Two of them
correspond to the so called strong promoters, while the one left is a
weak promoter \cite{weak_book}. Strong promoters show a high level of
expression in mRNA and usually their sequences are close to the
consensus sequence. The strong promoters studied here are the P5 virus
promoter, given by the 69 bp sequence shown in \cite{Kalosakas04} and
the human collagen type I $\alpha2$ chain, given by the 80 bp sequence
shown in \cite{Alexandrov_PLoSCB}. Finally, the weak promoter is the
lac operon regulatory region, whose 129 bp sequence has been taken
from \cite{Apostolaki11}.

In Fig. \ref{fig:dendos} (top) we show a detail of the free
energy dendrograms for each of the mentioned sequences. The basin
structure consists of a big set of low occupied (high energy) basins,
and a small set which gathers almost the whole trajectory (see below).
This small set of basins is the one shown in Fig. \ref{fig:dendos}.

In the botton of Fig. \ref{fig:dendos} some remarkable states for
each of the three promoters are highlighted.  The method 
identifies states with a biological meaning as they correspond to the
most important basins. The most significant sites we are dealing with are the transcription starting site (TSS) and the TATA box, although additional  promoter sequences can be found depending on the genome. The RNA polymerase binds to the TSS, starting the transcription into mRNA. Promoter sequences are usually labeled from the TSS ($+1$). The TATA-box is found approximately $35$ base pairs upstream from the TSS \cite{weak_book}. 

For the collagen chain, three states have been highlighted. State A
identifies the TSS, showing a bubble in this region with the particle
placed just there. States B and C are linked to excitations of other
important sites such as the TATA box (state B), see
\cite{Alexandrov_PLoSCB}. In the same way, we have found a basin
related to the TSS in the case of the P5 chain (state C) and the lac
operon (state C), together with other regulatory sites.

The arrangement of these basins in the free energy dendrogram informs
about the relative free energy between the states and the relation
between them. For example, the collagen dendrogram contains three main
branches, each one related to each of the three states shown. The P5
promoter shows an analogous structure: two main branches and another
one divided into two states (B and C) which are kinetically close. The
remaining states of each branch correspond to states similar to that
shown, with only slight variations in the chain conformation or in the
particle position.

\begin{figure}
\includegraphics[width=0.40\textwidth]{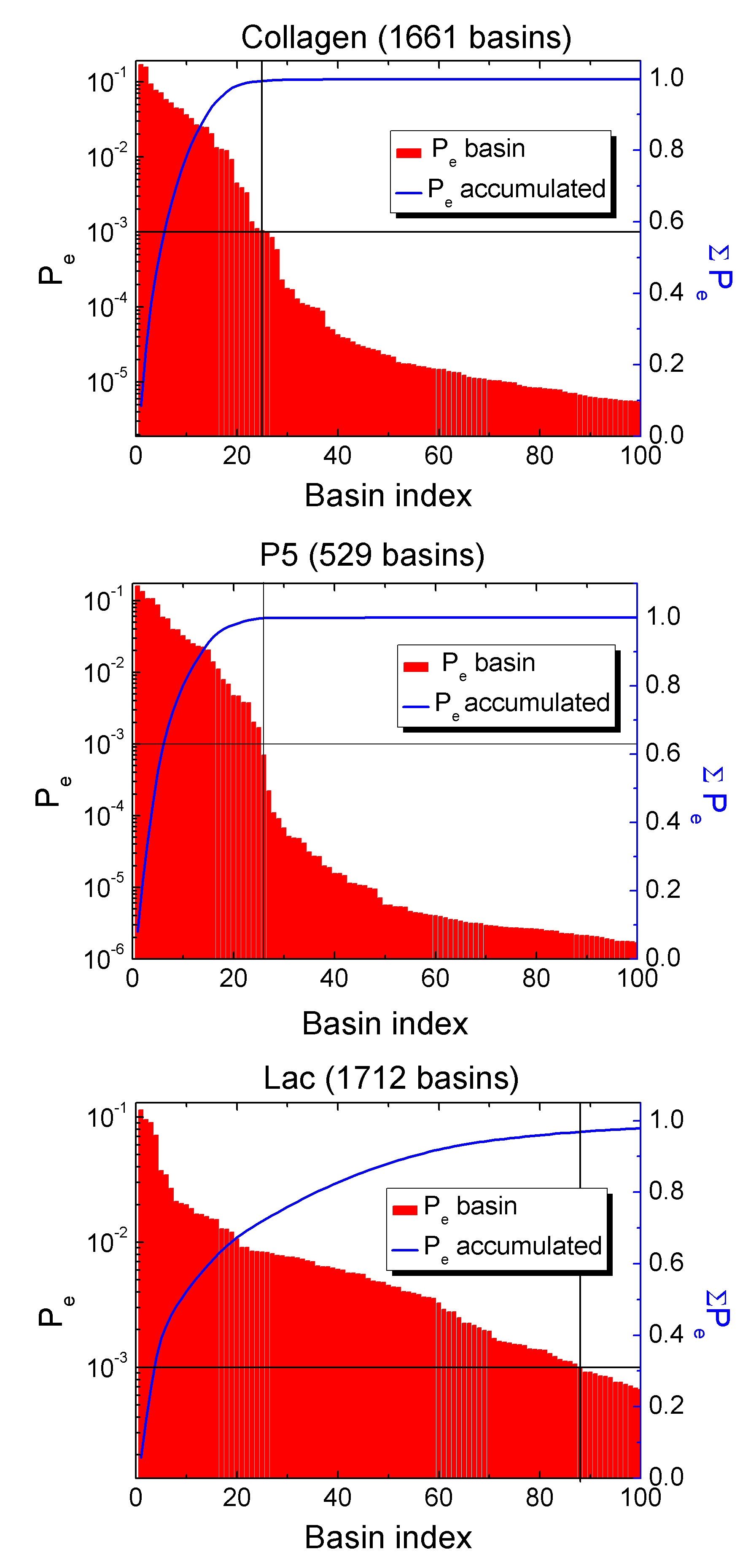}
\caption{(Color online) Basin occupancy (bars) together with accumulated weight (solid line) for the collagen, P5 and lac sequences. The
horizontal line shows the weight threshold between specific and nonspecific states. The vertical line establishes such a frontier in terms
of basins. 
Note the logarithmic y axis.}
\label{fig:ranks} 
\end{figure}

The lac promoter shows a clearly different behavior compared with the
two strong promoters. From a qualitative point of view, the
arrangement of basins differs from the P5 sequence or the collagen
one. To visualize quantitatively the difference, the basin occupancy
is plotted in Fig. \ref{fig:ranks}. We show the weight of each basin
(red bars) for the three sequences together with the accumulated
weight (blue line). It is remarkable, in the case of the collagen
sequence, that a few basins ($25$ out of $1661$) accumulate almost the
whole weight of the network (over the $99\%$). The results for the P5 promoter are completely analogous, a few basins account for most information of the dynamics.
These basins are the ones shown in the dendrogram of Fig. \ref{fig:dendos}. 
When we inspect the bottom graph in Fig. \ref{fig:ranks} we see a
completely different tendency. In the lac network, the distribution of
weight among the basins is more uniform. 

For the collagen and P5 networks we can define a
threshold from which the individual contribution to the total
weight is negligible. The trajectory is concentrated in around $25$ basins
and the remainder of the network can be seen as a ``background".  This
``background" is limited to those basins with a weight below
$10^{-3}$ (horizontal lines in Fig \ref{fig:ranks}).  Following this criterion,
we can distinguish between \emph{specific} and
\emph{nonspecific} states. Those basins above the threshold (vertical
lines in Fig. \ref{fig:ranks}) may be defined as specific states (with
a clear biological function) while those below the threshold may be
defined as non-specific states.

For the collagen sequence, $25$ ``specific'' basins appear, covering
$99.41\%$ of the total trajectory. The P5 and lac sequences show
respectively $23$ and $88$ ``specific'' basins, which gather $99.38\%$
and $96.91\%$ respectively of the total network weight. Using these
definitions, we are able to calculate the relative free energy between
states. These magnitudes reveal the ``strength'' of the different
sites in each promoter. It has been reported that specific binding
proteins show a greater affinity for strong promoters than for weak
ones \cite{Bintu05}. To quantify these differences we calculate
thermodynamical properties of the most important basins.  Once we have
divided the network into the different basins of attraction, several
statistical magnitudes can be defined from them. The weight of the
basin is defined as the sum of that of the nodes belonging to the
basin, i. e. for a basin $\alpha$ we have $P_\alpha=\sum_iP_i$ with
$i\in \alpha$. In the same way the entropy of each basin can be
defined as $S_\alpha/k_B=-\sum_iP_i\log P_i$ with
$i\in\alpha$. Attending to the previous definition of the nonspecific
basin, whose thermodynamical magnitudes can be computed as explained,
we can calculate the free energy of each basin with respect to the
nonspecific state. If $P_\beta$ is the weight of the non-specific
basin, then the free energy difference between a basin $\alpha$ and
the macrostate $\beta$ is $\Delta F_\alpha/k_BT=-\log(P_\alpha/
P_\beta)$. Table \ref{my_table} shows significant differences between
strong and weak promoters. On the one hand we observe that both the
total weight and entropy of the non-specific states in the weak
promoter exceed by almost an order of magnitude the ones shown for
the strong promoters. On the other hand, we can see that the specific
states show much higher free energy differences with respect to the
nonspecific states in the case of the strong promoters than the ones
shown for the lac sequence.  Thus, the analysis presented here
opens the way to a systematic study of promoter character within the framework of
a mesoscopic model.

\newcolumntype{C}{>{\centering\arraybackslash}X}
\newcolumntype{D}{>{\arraybackslash}X}
\begin{table}
\begin{center}
  \begin{tabularx}{0.49\textwidth}{C D C C D}
\hline \hline
   Promoter & State & $P_e\;$&$S/k_B\;$&$-\Delta F/kT$ \\ \hline
\multirow{5}{*}{Collagen} & A (TSS) & 0.169 & 1.365 & 3.305 \\ 
                   &B(TATA) & 0.157  & 1.380 & 3.232 \\ 
                   &C & 0.086 & 0.652 & 2.519 \\ 
                   &NS & 0.006 & 0.085 & 0.000 \\ \hline
\multirow{5}{*}{P5} & A(TATA) & 0.135 & 1.051 & 3.130 \\
                   & B & 0.107 & 0.913 & 2.898 \\ 
                   & C (TSS) & 0.086 & 0.684 & 2.681 \\ 
                   & D & 0.059 & 0.494   & 2.301 \\ 
                   & NS & 0.006 & 0.027 & 0.000 \\ \hline
\multirow{5}{*}{lac} & A(TATA) & 0.115 & 0.970 & 1.311 \\ 
                   & B & 0.095 & 0.891 & 1.120 \\ 
                   & C (TSS) & 0.090 & 0.775 & 1.066 \\ 
                   & D & 0.038 & 0.373 & 0.204 \\ 
                   & NS & 0.031 & 0.390 & 0.000\\ \hline \hline
   \end{tabularx}
\end{center}
\caption{Statistical (occupation probabilities) and thermodynamical
  (entropy and free energy differences) magnitudes calculated for the
  chosen states of Fig. \ref{fig:dendos} and the nonspecific states
  (NS), according to the criteria shown in Fig. \ref{fig:ranks}.}
  \label{my_table}
\end{table}

\begin{figure}
\includegraphics[width=0.40\textwidth]{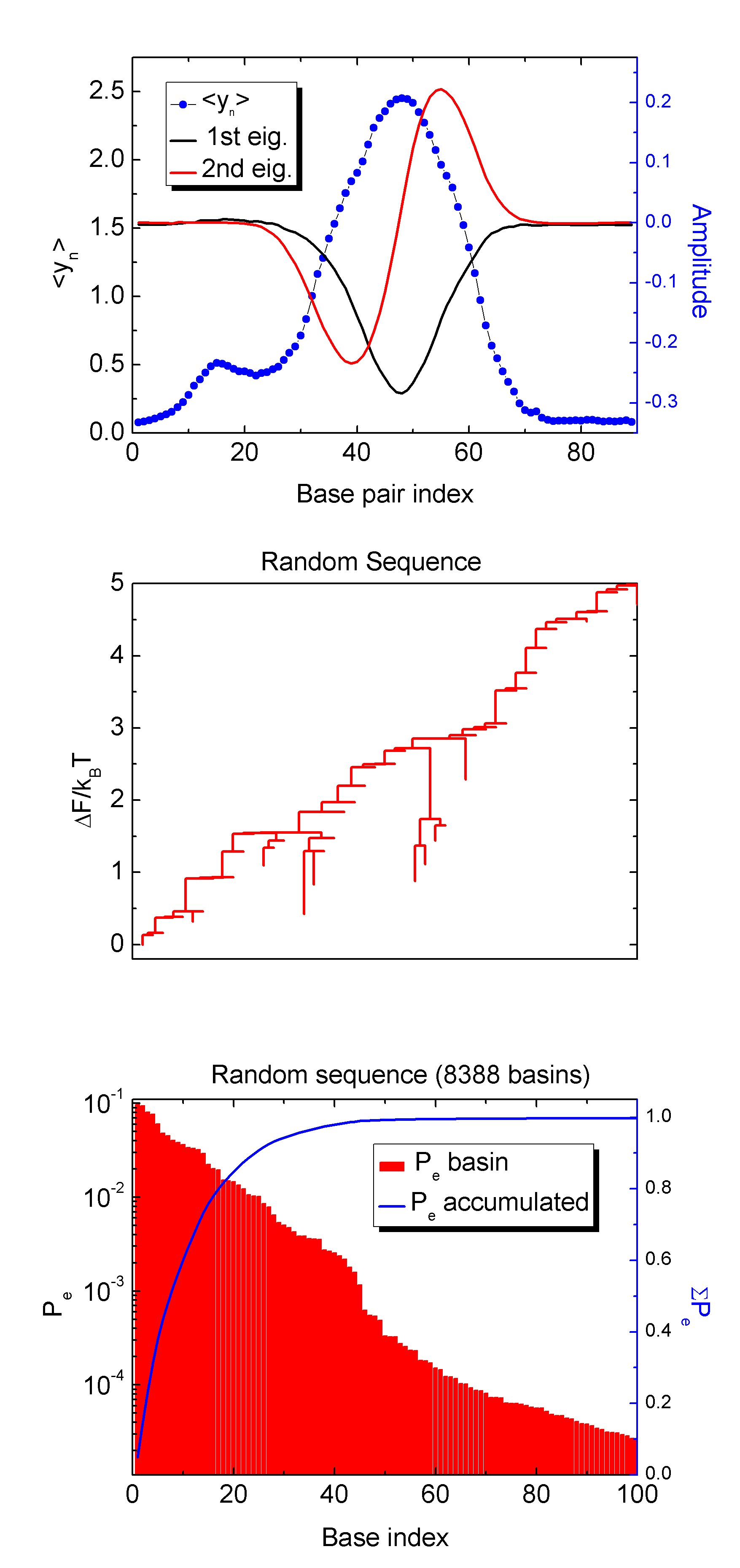}
\caption{(Color online) Analysis of a random sequence. Top: Probability of aperture
  along with the first two PCA eigenvectors. Middle: Free energy
  dendrogram. Bottom: Basin occupancy (bars) together with
  accumulated weight (solid line) }
\label{fig:random} 
\end{figure}

In addition to the three promoter sequences of real biological
systems, we have analyzed a random sequence in order to prove the
validity of our model. The random sequence has been obtained by taking
the P5 promoter sequence and shuffling its base pairs, so that the
obtained sequence contains the same base pairs but in random
positions. This sequence should contain no genetic information at all,
and this fact must be reflected in our analysis.

When analyzing the random sequence with our method, we can see huge
differences compared with the P5 promoter, as we would expect (see Fig.
\ref{fig:random}). First the structure of the network is completely
different. As there are no prominent states in the sequence, the
number of basins is huge (8388 compared with the 529 in the P5
promoter). The first two eigenvectors are representative of a
homogenous lattice without localized states. The distribution of
weights is also clearly different as now the ``background" basins keep
 $6 \%$ of the total network weight, an even  higher value than that
of the background basins gathered in the weak promoter. The dendrogram
also shows a much more distributed structure where, even though some
nodes appear to fall to privileged positions, their relevance within
the whole network structure is far from being comparable to that shown
in networks from biological promoters. All these facts validate our
model, as we can clearly distinguish between a sequence with binding
sites, and thus with biological information, and one with none, even
though their chemical composition is the same.

\section{Conclusions}

In this paper we have proposed and analyzed a mesoscopic model for the
characterization of binding sites on DNA promoter sequences. The model is
based on the 1D diffusion of an extended probe particle along the DNA chain.  
The particle is coupled to the opening states of the
chain (bubbles). In its dynamics, it visits the main sites
of the sequences, with dwelling times covering a
high percentage of the trajectory.  Such behavior has allowed us to
perform a deep analysis of the FEL which reveals the structure of the
complex phase space. The analyzed promoter sequences have been chosen
to include genomes from organisms of different domains (virus, bacteria
and eukaryote) and different strengths of expression.  The model and
the analysis used are able to capture the main biological details of
the sequences.

Our model gives energy differences between specific and nonspecific
sites of the promoter. Our results are in good {\em relative}
agreement with some data in the literature (see for instance
\cite{Bintu05}): they account for energy ratios between weak and
strong promoters. This fact would  also make possible the study of sequences 
in which several TSSs are involved, showing the relative strength between them.

We think that our results show the power of
coarse-grained or phenomenological mesoscopic models to qualitatively
and quantitatively analyze complex biological systems, in particular the
problem of protein-DNA regulatory and transcriptional interactions. 
Protein-DNA interaction is a fundamental problem which has been the
object of a very intense research from many different points of view
in the past years \cite{search, Berg}. 
Our system can be seen as the searching problem of a
universal protein on a given DNA sequence, providing an approach
for the study of specific protein-DNA interactions at the mesoscopic
level, where different protein will interact in different ways with
DNA molecules.


\begin{acknowledgments}
We thank Mar\'{\i}a F. Fillat for helpful discussions on promoter biology. 
The work is supported by the Spanish Projects No. FIS2008-01240 
and No. FIS2011-25167, cofinanced by Fondo
Europeo de Desarrollo Regional (FEDER) funds.
\end{acknowledgments}

\end{document}